\documentclass[%
 reprint,
 superscriptaddress,
 amsmath,amssymb,
 aps,
]{revtex4-2}

\usepackage{graphicx}
\usepackage{dcolumn}
\usepackage{bm}
\usepackage{hyperref}    
\hypersetup{
  colorlinks=true,
  allcolors=blue,
}
\usepackage{siunitx}

\begin{document}

\preprint{APS/123-QED}

\title{Strain Enhanced Spin Readout Contrast in Silicon Carbide Membranes}

\author{Haibo Hu}
 \altaffiliation{These authors contributed equally to this work.}
\affiliation{Ministry of Industry and Information Technology Key Lab of Micro-Nano Optoelectronic Information System, Guangdong Provincial Key Laboratory of Semiconductor Optoelectronic Materials and Intelligent Photonic Systems, Harbin Institute of Technology, Shenzhen 518055, P. R. China.}
\affiliation{Pengcheng Laboratory, Shenzhen 518055, P. R. China.}

\author{Guodong Bian}
 \altaffiliation{These authors contributed equally to this work.}
\affiliation{HUN-REN Wigner Research Centre for Physics, Institute for Solid State Physics and Optics, PO Box 49, H-1525, Budapest, Hungary.}

\author{Ailun Yi}
 \altaffiliation{These authors contributed equally to this work.}
\affiliation{State Key Laboratory of Materials for Integrated Circuits, Shanghai Institute of Microsystem and Information Technology, Chinese Academy of Sciences, Shanghai 200050, P. R. China.}
\affiliation{The Center of Materials Science and Optoelectronics Engineering, University of Chinese Academy of Sciences, Beijing 100049, P. R. China.}

\author{Chunhui Jiang}
\affiliation{Ministry of Industry and Information Technology Key Lab of Micro-Nano Optoelectronic Information System, Guangdong Provincial Key Laboratory of Semiconductor Optoelectronic Materials and Intelligent Photonic Systems, Harbin Institute of Technology, Shenzhen 518055, P. R. China.}

\author{Junhua Tan}
\affiliation{Ministry of Industry and Information Technology Key Lab of Micro-Nano Optoelectronic Information System, Guangdong Provincial Key Laboratory of Semiconductor Optoelectronic Materials and Intelligent Photonic Systems, Harbin Institute of Technology, Shenzhen 518055, P. R. China.}

\author{Qi Luo}
\affiliation{Ministry of Industry and Information Technology Key Lab of Micro-Nano Optoelectronic Information System, Guangdong Provincial Key Laboratory of Semiconductor Optoelectronic Materials and Intelligent Photonic Systems, Harbin Institute of Technology, Shenzhen 518055, P. R. China.}

\author{Bo Liang}
\affiliation{Ministry of Industry and Information Technology Key Lab of Micro-Nano Optoelectronic Information System, Guangdong Provincial Key Laboratory of Semiconductor Optoelectronic Materials and Intelligent Photonic Systems, Harbin Institute of Technology, Shenzhen 518055, P. R. China.}

\author{Zhengtong Liu}
\affiliation{Pengcheng Laboratory, Shenzhen 518055, P. R. China.}

\author{Xinfang Nie}
\affiliation{Department of Physics, State Key Laboratory of Quantum Functional Materials, and Guangdong Basic Research Center of Excellence for Quantum Science, Southern University of Science and Technology, Shenzhen 518055, China}
\affiliation{Quantum Science Center of Guangdong-HongKong-Macao Greater Bay Area (Guangdong), Shenzhen 518045, China.}

\author{Dawei Lu}
\affiliation{Department of Physics, State Key Laboratory of Quantum Functional Materials, and Guangdong Basic Research Center of Excellence for Quantum Science, Southern University of Science and Technology, Shenzhen 518055, China}
\affiliation{Quantum Science Center of Guangdong-HongKong-Macao Greater Bay Area (Guangdong), Shenzhen 518045, China.}

\author{Shumin Xiao}
\affiliation{Ministry of Industry and Information Technology Key Lab of Micro-Nano Optoelectronic Information System, Guangdong Provincial Key Laboratory of Semiconductor Optoelectronic Materials and Intelligent Photonic Systems, Harbin Institute of Technology, Shenzhen 518055, P. R. China.}
\affiliation{Pengcheng Laboratory, Shenzhen 518055, P. R. China.}
\affiliation{Quantum Science Center of Guangdong-HongKong-Macao Greater Bay Area (Guangdong), Shenzhen 518045, China.}
\affiliation{Collaborative Innovation Center of Extreme Optics, Shanxi University, Taiyuan 030006, Shanxi, P. R. China.}

\author{Xin Ou}
 \email{ouxin@mail.sim.ac.cn}
\affiliation{State Key Laboratory of Materials for Integrated Circuits, Shanghai Institute of Microsystem and Information Technology, Chinese Academy of Sciences, Shanghai 200050, P. R. China.}
\affiliation{The Center of Materials Science and Optoelectronics Engineering, University of Chinese Academy of Sciences, Beijing 100049, P. R. China.}

\author{Ádám Gali}
 \email{gali.adam@wigner.hun-ren.hu}
\affiliation{HUN-REN Wigner Research Centre for Physics, Institute for Solid State Physics and Optics, PO Box 49, H-1525, Budapest, Hungary.}
\affiliation{Department of Atomic Physics, Institute of Physics, Budapest University of Technology and Economics, M\H{u}egyetem rakpart 3., 1111 Budapest, Hungary.}
\affiliation{MTA-WFK Lendület “Momentum” Semiconductor Nanostructures Research Group, PO Box 49, H-1525, Budapest, Hungary.}

\author{Yu Zhou}
 \email{zhouyu2022@hit.edu.cn}
\affiliation{Ministry of Industry and Information Technology Key Lab of Micro-Nano Optoelectronic Information System, Guangdong Provincial Key Laboratory of Semiconductor Optoelectronic Materials and Intelligent Photonic Systems, Harbin Institute of Technology, Shenzhen 518055, P. R. China.}
\affiliation{Quantum Science Center of Guangdong-HongKong-Macao Greater Bay Area (Guangdong), Shenzhen 518045, China.}

\author{Qinghai Song}
 \email{qinghai.song@hit.edu.cn}
\affiliation{Ministry of Industry and Information Technology Key Lab of Micro-Nano Optoelectronic Information System, Guangdong Provincial Key Laboratory of Semiconductor Optoelectronic Materials and Intelligent Photonic Systems, Harbin Institute of Technology, Shenzhen 518055, P. R. China.}
\affiliation{Pengcheng Laboratory, Shenzhen 518055, P. R. China.}
\affiliation{Quantum Science Center of Guangdong-HongKong-Macao Greater Bay Area (Guangdong), Shenzhen 518045, China.}
\affiliation{Collaborative Innovation Center of Extreme Optics, Shanxi University, Taiyuan 030006, Shanxi, P. R. China.}



\date{\today}

\begin{abstract}
Quantum defects in solids have emerged as a transformative platform for advancing quantum technologies. A key requirement for these applications is achieving high-fidelity single-spin readout, particularly at room temperature for quantum biosensing. Here, we demonstrate through \textit{ab initio} simulations of a primary quantum defect in 4H silicon carbide that strain is an effective control parameter for significantly enhancing readout contrast. We validate this principle experimentally by inducing local strain in silicon carbide-on-insulator membranes, achieving a readout contrast exceeding $60\%$ while preserving the favorable coherence properties of single spins. Our findings establish strain engineering as a powerful and versatile strategy for optimizing coherent spin--photon interfaces in solid-state quantum systems. 
\end{abstract}

\maketitle

\emph{Introduction.--}Quantum defects in solid-state materials have emerged as promising platforms for advancing quantum technologies \cite{aharonovich2016solid, atature2018material, wolfowicz2021quantum, luo2023fabrication}. A central requirement for these applications is the ability to efficiently initialize, manipulate, and read out spin states with high fidelity \cite{barry2020sensitivity}. Among these processes, optical spin readout contrast--the relative difference in photon counts between distinct spin states--is essential for ensuring the efficiency and reliability of quantum operations \cite{rogers2014all, neumann2010single, hopper2018spin, lai2024single, gulka2021room}. Low optical spin readout contrast significantly limits sensitivity in quantum metrology \cite{barry2020sensitivity} and restricts the feasibility of single-shot readout \cite{delteil2014observation, zhang2021high, anderson2022five, lai2024single}. Although substantial progress has been made in improving optical readout contrast at cryogenic temperatures through spin-to-charge conversion or resonant excitations \cite{neumann2010single, aslam2013photo, zhang2021high, anderson2022five, lai2024single}, at room temperature, the maximum achievable optical spin readout contrast remains at much lower values \cite{hopper2018spin, jaskula2019improved, barry2020sensitivity, li2022room}. Recently, optical pulse sequences have been optimized to increase the optical spin readout contrast \cite{wirtitsch2023}; nevertheless, the maximum readout contrast is still fundamentally limited by the inherent properties of the color centers that govern spin-selective fluorescence and optical spin-polarization. This limitation underscores the urgent need for novel strategies to enhance the off-resonant optical spin readout contrast and drive advancements in quantum technologies. 

In principle, the spin readout contrast of isolated solid-state spin-defect systems depends on the critical rates in the optical spin-polarization loop \cite{bian2025theory} involving radiative and intersystem crossing (ISC) transitions. The relative strength between radiative and ISC processes primarily governs spin-dependent fluorescence emission, and the rates of ISC transitions are significantly influenced by variations in spin-state energy gaps and spin-orbit coupling (SOC) \cite{goldman2015state, thiering2017ab, thiering2018theory}. However, the feasible methods for modulating radiative and ISC rates to improve spin readout contrast remain largely unexplored. As a result, a significant gap exists in the experimental enhancement of optical spin readout contrast through emission and ISC engineering in solid-state quantum defect systems. 

In this Letter, we introduce a comprehensive theoretical framework for analyzing strain-mediated modulation of radiative and ISC rates. Using first-principles calculations, we show that appropriate strain fields can significantly enhance the optical spin readout contrast of divacancy qubits in 4H-silicon carbide (SiC). Experimentally, we harness longitudinal strain in 4H-SiC membranes and achieve a $60.6\%$ off-resonant optical spin readout contrast for divacancy qubits at room temperature. These results establish strain engineering as a powerful and broadly applicable strategy for optimizing coherent spin--photon interfaces in SiC and other wide-bandgap semiconductor platforms.

\begin{figure}
\includegraphics[width=0.9\linewidth]{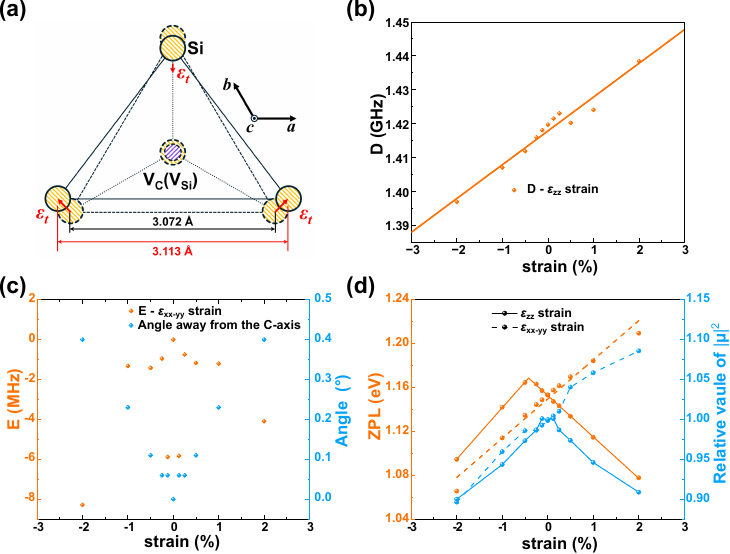}
\caption{\label{fig:calc_ZFS_ZPL}(a) Schematic diagram of the change of $hh$-divacancy due to the transverse strain $\varepsilon_{t}$ of ($xx-yy$). The \textit{c}-axis is perpendicular to the paper and extends outward. All yellow (purple) striped filled spheres represent Si (C) atoms. The dotted (solid) outline indicates the absence (presence) of strain. The black and red number with a unit of \AA ngstr\"om shows the distance between the two Si atoms neighboring the divacancy before and after applying the $1\%$ tensile transverse strain. (b) Calculated $D$ parameters under longitudinal strains. (c) Calculated $E$ parameters (orange) and the angle away from the \textit{c}-axis (blue) under transverse strains. (d) Calculated ZPL values (orange) and relative values of $|\bm{\mu}|^2$ (blue) under both longitudinal and transverse strains.}
\end{figure}

\emph{ZFS, ZPL, and readout contrast calculation.--}The effects of strain on the zero-field splitting (ZFS), zero-phonon line (ZPL), and optically detected magnetic resonance (ODMR) contrast in the well-studied \textit{hh}-divacancy center (the PL1 center~\cite{koehl2011}), which serves as a representative axial-divacancy in 4H-SiC, were systematically investigated using first-principles methods (see Supplementary Note~1 for details~\cite{supplementary}). Longitudinal ($\varepsilon_{zz}$) and transverse ($\varepsilon_{xx-yy}$) strains were applied to explore a broad parameter space. Tensile and compressive strains were introduced by directly elongating or contracting the lattice vectors to distort the supercell~\cite{udvarhelyi2018ab}. The longitudinal strain was applied along the \textit{c}-axis, preserving symmetry, while the transverse strain lowered the symmetry from $C_{3v}$ to $C_{1h}$ by distorting atomic positions. Fig.~\ref{fig:calc_ZFS_ZPL}(a) shows the lattice distortion induced by $1\%$ transverse strain, corresponding to a 0.041~\AA~ change in the distance between the nearest two Si atoms. 
The $\varepsilon_{zz}$ and $\varepsilon_{xx-yy}$ strains influence ZFS and ZPL in different ways. Since $\varepsilon_{zz}$ strain preserves $C_{3v}$ symmetry, the $E$ parameter (the anisotropic component in ZFS) remains zero, while the $D$ term (the axially symmetric component) is modulated. 

The calculated unstrained $D$ value is 1420~MHz, and a linear dependence of $D$ on $\varepsilon_{zz}$ strain was observed [Fig.~\ref{fig:calc_ZFS_ZPL}(b)], with a slope of $\SI{10 \pm 0.7}{\mega\hertz\per\text{(\% strain)}}$. In contrast, $\varepsilon_{xx-yy}$ strain induces minimal changes in $D$, varying by only 5~MHz across the $-1\%$ to $+1\%$ range. The angle of the $D$ tensor relative to the \textit{c}-axis peaks at just $0.4^\circ$ [Fig.~\ref{fig:calc_ZFS_ZPL}(c)], confirming the negligible influence of transverse strain on $D$. However, because transverse strain breaks the $C_{3v}$ symmetry, the $E$ term exhibits a parabolic response, reaching 8.3~MHz at $-2\%$ strain [Fig.~\ref{fig:calc_ZFS_ZPL}(c)]. 

A linear shift in ZPL energy is observed under $\varepsilon_{zz}$ strain [Fig.~\ref{fig:calc_ZFS_ZPL}(d)], consistent with recent results for nitrogen-vacancy (NV) centers in diamond~\cite{lopez2024quantum}. A linear coefficient of $\SI{35.7 \pm 1.9}{\milli\electronvolt\per\text{(\% strain)}}$ is extracted from the fit. Fig.~\ref{fig:calc_ZFS_ZPL}(d) also displays the effects of transverse strain on ZPL with a two-component behavior revealed: (i) a linear energy shift with strain magnitude and (ii) a peak centered at $-0.42~(\%~ \text{strain})$, exhibiting a 15.4~meV increase relative to the unstrained case. This asymmetric behavior is attributed to polarity-dependent Jahn--Teller distortions under tensile versus compressive transverse strain. In addition, the potential change under strains of radiative transition from $|^{3}E\rangle$ to $|^{3}A_{2}\rangle$ with a rate $k_{\mathrm{rad}}$ has also been demonstrated to be more comprehensive and is described as \cite{davidsson2020theoretical,bian2025theory}
\begin{equation}
k_{\mathrm{rad}} = \frac{n\left(2\pi\right)^3\nu^3\left|\bm{\mu}\right|^2}{3\epsilon_0hc^3},
\label{eq:k_rad}
\end{equation}
where $n$ = 2.647 is the refractive index for 4H-SiC~\cite{davidsson2020theoretical}, $\nu$ is the ZPL transition frequency, $\bm{\mu}$ is the transition dipole moment, $\epsilon_0$ is the vacuum permittivity, $h$ is the Planck constant, and $c$ is the speed of light in vacuum. From Eq.~\ref{eq:k_rad}, $k_{\mathrm{rad}}$ is proportional to both $|\bm{\mu}|^2$ and $\nu^3$. The relative values of $|\bm{\mu}|^2$ are shown in Fig.~\ref{fig:calc_ZFS_ZPL}(d), alongside the corresponding ZPL values (i.e., $\nu$).  

\begin{figure*} 
\includegraphics[width=0.9\linewidth]{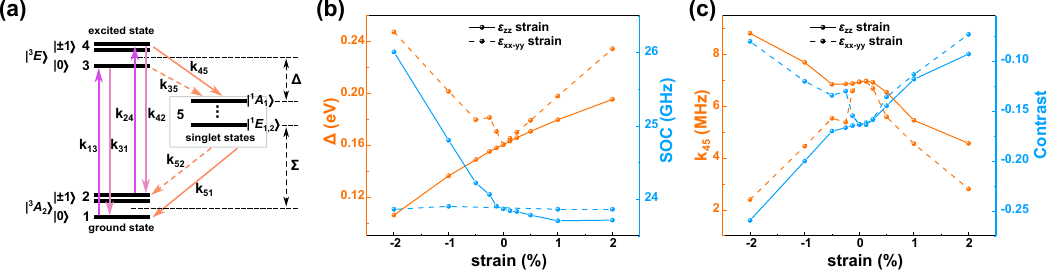}
\caption{\label{fig:calc_contr}(a) Five-level model with major associated rates and energy gaps. Pink arrows indicate the radiative transition and orange arrows indicate the strong (solid) and weak (dashed) nonradiative decay via the singlet states. $k_{ij}$ ($i, j = 1,2,3,4,5$ ) are rates for these transitions. (b) Calculated energy gap $\Delta$ and SOC results under various strains. (c) Calculated ISC transition rate $k_{45}$ and contrast under various strains.}
\end{figure*}

Next, the effect of strain on the spin readout contrast is analyzed. Fig.~\ref{fig:calc_contr}(a) displays our five-level rate-equation model with the major associated rates, which are categorized into non-radiative ISC transitions (with rates of $k_{35}$ and $k_{45}$) and radiative transitions (with rates of $k_{31}$ and $k_{42}$, where $k_{31} = k_{42} = k_{\mathrm{rad}}$). The contrast $C$ can be simplified and estimated by \cite{li2022room} 
\begin{equation}
C=\frac{\tau_{-1}-\tau_0}{\tau_0}=\frac{k_0-k_{-1}}{k_{-1}},
\label{eq:contr}
\end{equation}
with $k_0 = 1/\tau_0 = k_{31} + k_{35}$ (where $k_{35}$ is extremely weak to be ignored~\cite{li2022room}) and $k_{-1} = 1/\tau_{-1} = k_{42} + k_{45}$, where $\tau_0$ and $\tau_{-1}$ are the lifetimes of the excited states for $m_s = 0$ and $m_s = -1$, respectively. The ISC rate $k_{45}$ mainly depends on the energy gap $\Delta$ between triplet excited state $|^{3}E\rangle$ and singlet state $|^{1}A_{1}\rangle$, and the SOC term \cite{goldman2015state,thiering2017ab,thiering2018theory}, which may be regulated by strains. Fig.~\ref{fig:calc_contr}(b) displays the calculated $\Delta$ and SOC results under various strains. As the longitudinal strain increases (from negative to positive), the $\Delta$ values show an approximately linear trend of change, with the value decreasing, while the SOC values show an exponential downward trend and seem to be converging. Under increasing the transverse strain, the $\Delta$ first decreases and then increases, showing a V-shaped trend, while SOC remains almost unchanged. Mapping these calculated $\Delta$ and SOC, all the $k_{45}$ results are addressed and shown in Fig.~\ref{fig:calc_contr}(c). Finally, the effects of longitudinal and transverse strains on the contrast are calculated and presented in Fig.~\ref{fig:calc_contr}(c). Longitudinal strain primarily drives the enhancement of the contrast under lattice compression. The contrast becomes larger under the negative longitudinal strain of $-2\%$ and reaches a value of $-25.9\%$, which finally leads to a $59\%$ increase from the results without strain.

\begin{figure*} 
\includegraphics[width=0.9\linewidth]{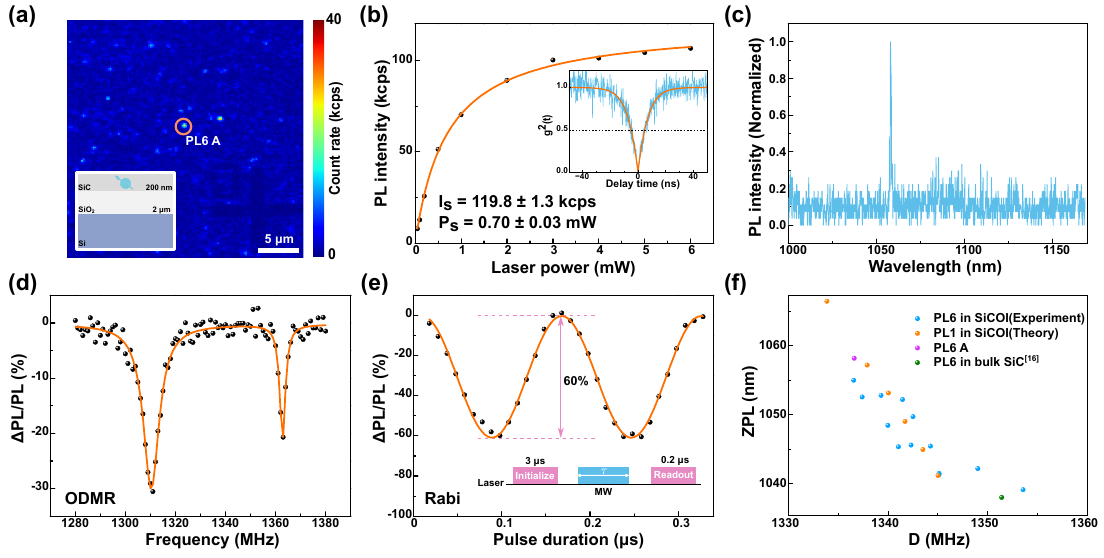}
\caption{\label{fig:expt_rabi_ODMR}Spin and optical properties of the single spin in SiCOI. (a) Confocal scanning image of the defect, with the inset showing a diagram of the cross-sectional view of the SiCOI sample. The SiC membrane has a thickness of 200~nm. (b) Saturation behavior. The black dots are the background-corrected experimental data, and the solid line is a fit using the function $I_P=I_s\cdot P(P+P_s)$, where $P$ and $I_P$ are the power of the excitation laser and the corresponding count rate, respectively. Insert is the second-order correlation function $g^2(t)$ measurement of the single spin. At an excitation laser power of 0.2~mW, $g^2(0)$ is well below 0.5, indicating single-photon emission. (c) Photoluminescence (PL) spectra of the single spin at 4~K reveal a ZPL peak near 1058.2~nm. (d) ODMR spectra of the single spin at zero magnetic fields show two prominent peaks. The black dots are the raw data, while the solid lines show the corresponding Lorentzian fits. (e) Rabi oscillations of the left peak in (d), the raw data was fitted with a cosine function, and the Rabi readout contrast is approximately $\SI{60.6 \pm 1.0}{\percent}$. (f) Comparison of the relationship between ZPL and $D$ for some strained defects with theoretical calculations.}
\end{figure*}

\emph{Spin and optical characterization of a single spin in SiCOI.--}Based on the above analysis, we have chosen SiCOI as the experimental framework for this study. This selection is driven by the fact that the SiCOI fabrication process inherently introduces longitudinal and transverse strain, which is challenging to achieve in bulk SiC \cite{falk2014electrically}. As shown in Fig.~\ref{fig:expt_rabi_ODMR}(a), we began by characterizing a single divacancy spin, as labeled as PL6~A, in a 200-nm thick SiCOI membrane. The membrane was fabricated via thinning and polishing technique \cite{lukin20204h} [Fig.~\ref{fig:expt_rabi_ODMR}(a) inset], avoiding ion-induced damage from the smart-cut method. A 4H-SiC wafer with an epitaxial SiO$_2$ layer was bonded to an oxidized Si wafer. This was followed by mechanical grinding and chemical mechanical polishing (CMP) to obtain a micrometer-thick layer~\cite{hu2024room}. The final thickness of 200~nm was achieved by dry etching the SiC layer(more details in Supplementary Note 2~\cite{supplementary}). This method inherently introduces significant interfacial strain--a combination of longitudinal ($\varepsilon_{zz}$) and transverse ($\varepsilon_{xx-yy}$) strain components--due to the lattice mismatch between SiC and SiO$_2$. Fig.~\ref{fig:expt_rabi_ODMR}(a) displays the confocal scanning map of the single defect PL6~A in the membrane.

The saturation curve of the single defect was measured under 914-nm continuous wave (CW) excitation, as shown in Fig.~\ref{fig:expt_rabi_ODMR}(b). The experimental data was fitted using the equation $I_P=I_s\cdot P(P+P_s)$, where $P$ and $I_P$ represent the power of the excitation laser and the corresponding photon count rate, respectively. The saturation power $P_s$ was determined to be \SI{0.70 \pm 0.03}{\milli\watt}, and saturation intensity $I_s$ was around $\SI{119.8 \pm 1.3}{\text{kcps}}$. This brightness is comparable to the PL6 divacancy center found in bulk SiC \cite{li2022room}. The second-order correlation function $g^2 (t)$, measured at an excitation power of 0.2~mW, is shown in the insert figure of Fig.~\ref{fig:expt_rabi_ODMR}(b). Background-correction with the formula $g^2\left(\tau\right)=\left[g_{raw}^2\left(\tau\right)-\left(1-\rho^2\right)\right]/\rho^2$ was performed where $\rho={s}/(s+b)$ and $s$ and $b$ are the signal and background counts, respectively \cite{wang2018bright}. The value of $g^2 (0)$ is well below 0.5, confirming the single photon emission \cite{falk2013polytype, li2022room}. Fig.~\ref{fig:expt_rabi_ODMR}(c) shows the PL spectrum of the single defect measured at 4~K. The ZPL was measured to be approximately 1058.2~nm. The ODMR measurements under \textit{c}-axis magnetic fields further corroborated this strain effect. As shown in Fig.~\ref{fig:expt_rabi_ODMR}(d), the ODMR spectrum exhibits two prominent peaks. Based on the measurements conducted by applying different magnetic fields during the ODMR measurements (details in Supplementary Note 4~\cite{supplementary}), the defect exhibits behavior characteristic of an axial divacancy. The spin Hamiltonian under strain \cite{wang2022zero} is
\begin{equation}
H=\left(D_{gs}+\ \Pi_z\right)S_z^2+\ \Pi_x\left(S_yS_y-\ S_xS_x\right)+\gamma_eB_zS_z,
\label{eq:Hamiltonian}
\end{equation}
where $D_{gs}$ is the ground state ZFS, $\Pi_{x/z}$ is the strain along the $x/z$ axis, $S$ is the electronic spin, $\gamma_e$ is the gyromagnetic ratio of the electronic spin, and $B_z$ is the $z$-component of the external magnetic field. From the fitting with the total Hamiltonian, as displayed in Fig. S1(a) \cite{supplementary}, the ZFS parameters $D = \SI{1336.6 \pm 0.1}{\mega\hertz}$ and $E = \SI{22.1 \pm 0.2}{\mega\hertz}$ were obtained. The left ODMR peak displays a contrast of approximately $30\%$, consistent with the characteristics of PL6 centers in bulk SiC \cite{li2022room}. Next, Rabi oscillations were measured by adjusting the duration of the microwave pulse between spin initialization and readout [Fig.~\ref{fig:expt_rabi_ODMR}(e)]. The raw data were fitted with a cosine function, revealing a Rabi contrast of around $\SI{60.6 \pm 1.0}{\percent}$. This signifies a twofold improvement over the contrast reported for PL6 divacancy spins in bulk SiC \cite{li2022room}. Furthermore, as shown in Fig.~\ref{fig:expt_rabi_ODMR}(f), a greater number of single PL6 spins in SiCOI were measured.

As the exact microscopic origin of the PL6 divacancy is yet to be confirmed, we use the calculated PL1 data for comparison with the observed strain-dependent parameters. Since the PL6 and PL1 divacancy spins share the same symmetry and spin Hamiltonian form, we assume that the strain-induced trends can be well captured using this method. To facilitate this comparison, the calculated absolute values of the $D$ and ZPL parameters were shifted by $-26$~MHz and $-78$~nm, respectively, to align them with the observed values at zero strain. The relationship between the $D$ and ZPL parameters as a function of strain is consistent with the theoretical calculations presented in Figs.~\ref{fig:calc_ZFS_ZPL}(b) and (d). This excellent agreement reinforces the interpretation of the experimental results: the observed extremely high optical spin readout contrast can be well explained by the large strain field present in SiCOI.

\begin{figure*}
\includegraphics[width=0.9\linewidth]{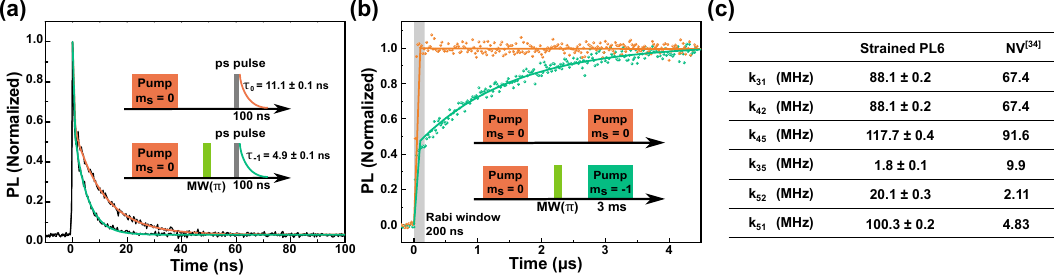}
\caption{\label{fig:expt_fl}(a) Spin-resolved excited state lifetime measurements for the single spin at $m_s$ = 0 or $m_s= -1$ at room temperature. The orange and green lines are the double exponential fitting, revealing a longer lifetime of $\SI{11.1 \pm 0.1} {\nano\second}$ and $\SI{4.9 \pm 0.1}{\nano\second}$ for $m_s = 0$ and $m_s= -1$, respectively, and a shorter decay from the system response or background fluorescence. (b) The fitting of the fluorescence traces for $m_s$ = 0 and $m_s= -1$ with the model in Fig.~\ref{fig:calc_contr}(a). (c) Comparison of the parameters fitted by the level model for the strained singe PL6 spin with diamond NV center in reference \cite{gupta2016efficient}.}
\end{figure*}

\emph{Temporal dynamics and models of the spin readout process.--}To uncover the underlying spin dynamics, spin-resolved excited-state lifetime measurements were conducted at room temperature for the $m_s$ = 0 and $m_s= -1$ states, as shown in Fig.~\ref{fig:expt_fl}(a). The orange and green lines illustrate the fitting with a double-exponential decay, in which the longer time parameters are lifetimes of $\tau_0 = \SI{11.1 \pm 0.1}{\nano\second}$ and $\tau_{-1} = \SI{4.9 \pm 0.1}{\nano\second}$ for $m_s = 0$ and $m_s= -1$, respectively, and the other fast decay may originate from the system response or background fluorescence decay \cite{christle2017isolated}. These lifetimes offer valuable insights into the spin-state dynamics and excited-state decay processes, affecting the overall spin readout contrast. The $C$ value of the strained PL6 is calculated to be $56\%$ using Eq.~\ref{eq:contr}, compared with the value of $33.6\%$ of PL6 in bulk SiC \cite{li2022room}. This value is close to the Rabi contrast of $60.6\%$, which we obtained experimentally in Fig.~\ref{fig:expt_rabi_ODMR}(e). Furthermore, we measured additional single-color centers (PL6~B-PL6~E) (Supplementary Note 4 \cite{supplementary}) and observed a Rabi contrast ranging from $21\%$ to $43\%$. Notably, as the Rabi contrast increased, the ratio of $C=(\tau_{-1}-\tau_0)/\tau_0$ also decreased, and these two values are close in magnitude [Fig. S3 \cite{supplementary}], consistent with the above analysis.

Besides direct lifetime measurement, we further explore the dynamics of the spin readout by performing time-resolved fluorescence measurements, as depicted in Fig.~\ref{fig:expt_fl}(b). The fluorescence traces following optical initialization ($m_s = 0$) are compared with and without applying a microwave $\pi$ pulse to flip the spin state. The gray-shaded region indicates the Rabi window, corresponding to the photon accumulation time ($\tau$ = 200~ns) used for Rabi oscillation in Fig.~\ref{fig:expt_rabi_ODMR}(e). The time window's duration is crucial for maximizing the contrast of the spin readout. A shorter time window yields fewer photons per cycle, requiring a balance between time resolution and photon count. For diamond NV centers, the optimal duration for this window is approximately 220~ns \cite{doherty2013nitrogen, hopper2018spin}. Another factor affecting the Rabi contrast is the excitation power. An increase in applied power leads to a faster spin initialization, which in turn results in a decrease in Rabi contrast, as detailed in Supplementary Note 4 \cite{supplementary}.

The time-resolved fluorescence dynamics were quantitatively analyzed using our five-level rate equation model [Fig.~\ref{fig:calc_contr}(a)], with the spin-dependent lifetimes from Fig.~\ref{fig:expt_fl}(a) serving as critical constraints, as detailed in Supplementary Note 3 \cite{supplementary}. All fitting parameters obtained are compared with those from the well-studied NV center in diamond \cite{gupta2016efficient}, as shown in Fig.~\ref{fig:expt_fl}(c). A prominent feature emerges in the strained PL6 system, where the $k_{45}$/$k_{35}$ ratio demonstrates a 7-fold enhancement compared to the NV center, directly mirroring our first-principles predictions of strain-reduced energy gaps $\Delta$ between the triplet excited state $|^{3}E\rangle$ and singlet ground state $|^{1}A_{1}\rangle$ \cite{bian2025theory}. This quantitative agreement bridges microscopic strain effects with macroscopic contrast improvement, where the $60.6\%$ Rabi contrast [Fig.~\ref{fig:expt_rabi_ODMR}(e)] emerges as a direct consequence of accelerated ISC transitions under compressive strain, as evidenced by the synergy between theoretical modeling and experimental observations.

\emph{Discussion and Outlook.--}Our combined theoretical and experimental study demonstrates that strain in the SiC membrane effectively modulates ISC transition rates, with axial strain playing a key role in enhancing non-radiative ISC pathways. Additionally, we experimentally achieve a spin readout contrast exceeding $60\%$ in strained SiCOI membranes. This approach is not only applicable to SiCOI but also extends to other solid-state defect systems, such as diamond membranes and 2D materials \cite{jing2024scalable, guo2024direct}. These results position strain engineering as a scalable method for optimizing spin readout contrast, providing a pathway for improved spin-photon interfaces in solid-state quantum systems.

We acknowledge the support from the  National Key R$\&$D Program of China (Grant No.\ 2021YFA1400802, 2022YFA1404601, 2023YFB2806700),the Innovation Program for Quantum Science and Technology (Grant No.\ 2024ZD0302100), the National Natural Science Foundation of China (Grant No.\ 12304568, 11934012, 62293520, 62293522, 62293521, 12074400 and 62205363), the GuangDong Basic and Applied Basic Research Foundation (Grant No.\ 2022A1515110382), Shenzhen Fundamental research project (Grant No.\ JCYJ20241202123903005, JCYJ20230807094408018), Guangdong Provincial Quantum Science Strategic Initiative (GDZX2403004, 2303001, GDZX2306002, GDZX2200001, GDZX2406002). A.G.\ acknowledges the support by the Ministry of Culture and Innovation and the National Research, Development and Innovation Office within the Quantum Information National Laboratory of Hungary (Grant No.\ 2022-2.1.1-NL-2022-00004) is much appreciated. A.G.\ acknowledges the high-performance computational resources provided by KIF\"U (Governmental Agency for IT Development) institute of Hungary and the European Commission for the projects QuMicro (Grant No.\ 101046911), SPINUS (Grant No.\ 101135699), and QuSPARC (Grant No.\ 101186889). Y.Z.\ acknowledges the support from Young Elite Scientists Sponsorship Program by CAST, Q.S.\ acknowledges New Cornerstone Science Foundation through the XPLORER PRIZE, A.Y.\ acknowledges the support from Shanghai Science and Technology Innovation Action Plan Program (Grant No.\ 22JC1403300), CAS Project for Young Scientists in Basic Research (Grant No.\ YSBR-69), the Major Key Project of PCL, the Talent Program of Guangdong Province (Grant No.\ 2021CX02X465).

\bibliography{apssamp}

\end{document}